# SHEARED FLOW AS A STABILIZING MECHANISM IN ASTROPHYSICAL JETS

Lucas F. Wanex and Erik Tendeland

*University of Nevada, Reno*

**Abstract**

It has been hypothesized that the sustained narrowness observed in the asymptotic cylindrical region of bipolar outflows from Young Stellar Objects (YSO) indicates that these jets are magnetically collimated. The $j_z \times B_\phi$ force observed in z-pinch plasmas is a possible explanation for these observations. However, z-pinch plasmas are subject to current driven instabilities (CDI). The interest in using z-pinches for controlled nuclear fusion has lead to an extensive theory of the stability of magnetically confined plasmas. Analytical, numerical, and experimental evidence from this field suggest that sheared flow in magnetized plasmas can reduce the growth rates of the sausage and kink instabilities. Here we propose the hypothesis that sheared helical flow can exert a similar stabilizing influence on CDI in YSO jets.

**Keywords:** astrophysical jets, linear analysis, sheared flow

# 1. Collimation of Astrophysical Jets

The sustained narrowness of the asymptotic cylindrical region of many YSO bipolar outflows spawns the hypothesis that an intrinsic collimating mechanism is present in the jet plasma. Here we use the term "asymptotic cylindrical region" to mean the narrow cone between the disk-wind initial compression region and the jet termination region in the ambient medium. A typical example of this comes from Hubble Space Telescope Observations of HH30 (Burrows et al. 1996). The apparent opening angle of this jet becomes narrower in the asymptotic cylindrical region of the outflow when compared to the opening angle near the jet source (Mundt et al. 1990). Similar examples of recollimation in other jets have also been observed (Königle and Pudritz 2000). Observations of this nature have lead to the concept that the jet narrowness is maintained by self-collimation (Shu et al. 2000). Self-collimation can be caused by hoop stresses from a toroidal magnetic field in the jet plasma.

This collimating mechanism is similar to the $j_z \times B_\phi$ force observed in z-pinch plasmas. If this mechanism is to be considered as the cause of the sustained narrowness of some YSO jets the possibility that these jets carry current far from the accretion disk must be admitted. It is well known that a cylindrical current-carrying plasma column with a helical magnetic configuration is subject to MHD instabilities (Bateman 1978; Chandrasekhar 1961). The key instabilities are the interchange (sausage) and kink modes (Freidberg 1987). The sausage *m* = 0 and kink *m* = 1 instabilities have been invoked to explain the observation of knots, wiggles, and filamentary structures in astrophysical jets (Nakamura and Meier 2004, Reipurth and Heathcote 1997). It has also been suggested that the disruption caused by the kink instability discredits the magnetic collimation model (Spruit et al. 1997). However, it is an open question whether MHD instabilities will disrupt YSO jet collimation (Königle and Pudritz 2000).

Two mechanisms for reducing z-pinch plasma instabilities may explain why this is so. Z-pinch plasmas will be stable if the ratio of axial to azimuthal magnetic field strength is greater than the Kruskal-Shafranov limit (Kruskal and Schwarzschild 1956; Shafranov 1954). Some stellar evolution models predict that low-mass protostars form in an interstellar medium that is supported by a magnetic field (Ray 2004; Königl and Pudritz, 2000). This interstellar field could supply the instability reducing mechanism in these jets; however the efficiency of the pinch effect is reduced because the axial magnetic field must be compressed as well as the jet plasma itself (Shumlak and Hartman 1995). Analytical, experimental and numerical results show that sheared flow in z-pinch plasmas can reduce the growth of MHD instabilities (Arber and Howell 1995, Bateman 1978, DeSouza-Machado et al. 2000, Golingo et al. 2005, Shumlak and Hartman 1995, Sotnikov et al. 2002, Sotnikov et al. 2004, Ruden 2002, Winterberg 1985, Winterberg 1999, Wanex et al. 2004, Wanex 2005a). Velocity shear stabilization is primarily a phase mixing process that disrupts the growth of unstable modes (DeSouza-Machado et al. 2000, Wanex et al. 2005b).

There are sound theoretical reasons why Keplerian shear could be present in jets originating from an accretion disk (Bacciotti et al. 2004, Bally et al. 2002, Völker et al. 1999). The disk-wind model of jet formation in protostars postulates that the source of the jet comes from a wide radial band of the accretion disk (Königl and Pudritz 2000). Since the rotational motion in the disk varies with distance from the central object the disk-wind driven jet velocity profile may also vary with distance from the central object. Axial and rotational sheared flows have been observed in astrophysical jets (Bally et al. 2002, Bacciotti et al. 2002, Bacciotti et al. 2003, Bacciotti et al. 2004, Coffey et al. 2004).

In this paper we will focus on the stabilizing effects of sheared helical flow. Due to space limitations a brief summary of the evidence that sheared flow reduces instabilities in z-pinch plasmas will be presented and the possibility that sheared flow could be responsible for the sustained narrowness in YSO jets will be considered.

## 2. MHD Model of an Astrophysical Jet

The MHD jet is modeled as perfectly conducting cylindrical plasma. The jet is then considered as a region in space where supersonic plasma and an electric current flow. The electric current *I* is balanced by a return current *Ir* of equal size. This return current can be modeled as a diffuse flow in the ambient medium or as a sheet on the jet surface (Lery and Frank 2000). Jet models with return current have been referred to as "cocoon jets" (Appl and Camenzind 1992, Lesch et al. 1989, Nakamura and Meier 2004). MHD instabilities grow, but do not propagate in stationary z-pinch plasmas (Appl et al. 2000). The kink mode in a plasma column moving with uniform velocity would simply move with the flow and thus grow at the same rate as a stationary kink (Shumlak and Hartman 1995). By transforming to a frame that moves with the jet, the jet plasma becomes analogous to a stationary cylindrical z-pinch plasma. The velocity of the jet in this frame is modeled with

$$\mathbf{v} = \left[ 0 \ , \ v_{0\phi}/\sqrt{a+r} \ , \ v_{0z}\left(1/\sqrt{a} - 1/\sqrt{a+r}\right)\right] \tag{1}$$

(in cylindrical coordinates) where $v_{0\phi}$ and $v_{0z}$ are constants of proportionality and *a* is a small number to prevent the velocity from going to infinity at $r=0$. The functional form of both velocity components is approximately Keplerian. In a frame that moves with the axial velocity at the center of the jet the axial plasma flow appears to be increasingly swept back with increasing radius. At the origin the axial velocity is zero and increases to a maximum at the edge of the jet. Both the azimuthal and axial components of velocity have Keplerian shear because the disk-wind is modeled as the source of plasma in the jet. This model is intended to simulate the asymptotic jet cylinder far from the surface of the accretion disk and termination region.

### 3. Stability Criteria

#### 3.1 Kelvin-Helmholtz instability

The form of the Kelvin-Helmholtz instability (KHI) considered here occurs at the tangential boundary between the edge of the jet and the ambient medium. The KHI can be excited by the velocity discontinuity that exists at this boundary (Keppens et al. 2005). For astrophysical jets this instability will be excited if the velocity discontinuity is greater than the Alfvén velocity (Nakamura and Meier 2004). For the Keplerian velocity profile considered here the velocity at the outer edge of the jet is lower than the velocity at the center of the jet (relative to the ambient medium). Sub-Alfvénic discontinuities are in general not subject to KHI so our analysis will focus on velocity profiles that meet this stability criterion.

#### 3.2 Sausage instability

The sausage instability is an interchange mode that can cause axisymmetric pinches or bulges to grow exponentially in the jet plasma. It has been shown analytically that the sausage instability can be stabilized in z-pinch plasmas with sheared axial flow $V'$ meeting the following criteria

$$V' > \gamma \sqrt{\ln(R)}, \qquad (2)$$

where $\gamma$ is the growth rate and $R$ is a dimensionless parameter analogous to the Reynolds number (DeSouza-Machado et al. 2000). In this case $\gamma \sim v_T / r_0$ where $v_T$ is the average thermal velocity and $r_0$ is the radius of the plasma. If we let $V' \sim \nabla V / r_0$, where $\nabla V$ is the difference in the axial velocity between the center of the jet and the edge of the jet, the stability requirement is

$$\nabla V > \sqrt{\ln(R)}\, \mathrm{v}_T. \tag{3}$$

Thus we see that $\nabla V$ must be above a threshold to prevent the sausage instability.

### 3.3 Kink instability

A global stability analysis of the kink mode using non-relativistic compressible ideal MHD with gravity neglected will demonstrate the stabilizing influence of sheared flow in current-carrying jets. The MHD equations are made dimensionless by normalizing the scales to the average radius of the jet in the asymptotic cylindrical region, the average thermal velocity of the jet plasma, the ambient pressure, and the ambient density. The linearized equations are solved numerically with a generic two-step predictor-corrector, second-order accurate space and time-centered advancement scheme (see Wanex et al. 2005b for details). The problem is treated by introducing perturbations into the plasma equilibrium state and following their linear development in time. All perturbed plasma variables (magnetic field, density, pressure, and velocity) have the form $\xi(r)e^{i(k_z z + \theta - \omega t)}$. Initially the growth rates of the perturbed plasma variables are uncorrelated, however, after several growth times the solution converges to the fastest growing unstable mode. For this analysis the jet is considered to be in equilibrium with the surrounding medium across its boundary at $r = r_0$. The use of fixed boundary conditions allows a global stability analysis of internal unstable modes (Arber and Howell 1995, Appl et al. 2000).

It has been shown that sheared azimuthal flow is effective at reducing the growth rate of the kink instability in z-pinch plasmas but has little effect on axisymmetric modes (see Wanex et al. 2005b for a detailed explanation). Sheared axial flow is effective for stabilizing the sausage mode (DeSouza-Machado et al. 2000). This suggests that the growth of both the sausage and kink instabilities can be reduced by combining axial and azimuthal velocity components to produce helical sheared flow. For this reason sheared helical flow will be used in this analysis of the kink instability.

Here we present the results of this analysis for two equilibrium profiles. The parabolic profile is obtained by using the magnetic field produced by a constant current density in the jet. We also present results for the constant electron velocity (Bennett) equilibrium profile with field maximum at $2r_0/3$.

## 5. Results

Figure 1 shows the results of the linear analysis on the kink instability for the parabolic equilibrium profile with azimuthal velocity $0.3/\sqrt{0.1+r}$. The instability growth rates are reduced to zero for $v_{0z} > 1$. Figure 2 shows the results for the Bennett equilibrium profile with the same velocity as in figure 1. The instability growth rates are reduced to zero for $v_{0z} > 1.2$. This can be interpreted to mean that sheared helical flow can stabilize the kink instability for the parameters and profiles considered here if

$$v_{0z} > 1.2. \tag{4}$$

Using (1) and (3) one finds that

$$v_{0z} > \frac{\sqrt{\ln(R)}}{2.2} v_T \tag{5}$$

is the stability criterion for the sausage instability ($r=1$ and $a=0.1$). Thus if (4) and (5) are satisfied the growth rates for both the sausage and kink instability can be reduced to zero for both of these examples. The Kelvin-Helmholtz stability condition can also be satisfied if the velocity at the edge of the jet is sub-Alfvénic in a frame at rest with respect to the ambient medium.

## 6. Conclusion

The results of this analysis suggest that the Kelvin-Helmholtz, sausage and kink instabilities in current carrying jets can be suppressed by Keplerian helical sheared flow for some equilibrium profiles. These results are sufficiently positive to motivate further analysis of the hypothesis that sheared helical flow can stabilize YSO jets. More work is required to extend the investigation to a larger range of parameters and equilibrium profiles.

# References

Appl, S. and Camenzind M.: 1992, *Astron. Astrophys.* **256**, 354.

Appl, S., Lery, T. and Baty, H.: 2000, *Astron. Astrophys.* **355**, 818.

Arber, T. D., and Howell, D. F.: 1995, *Phys. Plasmas* **3**, 554.

Bacciotti, F., Ray, T. P., Mundt, R., Eislöffel, J., and Solf, J.: 2002, *ApJ* **576**, 222.

Bacciotti, F., Ray, T. P., Eislöffel, J., Woitas, J., Solf, J., Mundt, R., and Davis, C. J.: 2003, *Astrophys. Space Sci.* **287**, 3.

Bacciotti, F., Ray, T. P., Coffey, D., Eislöffel, J., and Woitas, J.: 2004, *Astrophys. Space Sci.* **292**, 651.

Bally, J., Heathcote, S., Reipurth, B., Morse, J., Hartigan, P. and Schwartz, R.: 2002, *Astron. J.* **123**, 2627.

Bateman, G., Schneider, W. and Grossman, W.: 1974, *Nucl. Fusion* **14**, 669-683.

Bateman, G.: 1978, *MHD Instabilities*, (The MIT Press, Cambridge).

Burrows, C. J., Stapelefeldt, K. R., Watson, A., M., et al.: 1996 ApJ **473**, 437.

Chandrasekhar, S.: 1961, *Hydrodynamic And Hydromagnetic Stability*, (Clarendon Press, Oxford).

Coffey, D., Bacciotti, F., Woitas, J., Ray, T. P. and Eislöffel, J.: 2004, Ap&SS **292**, 553.

Coker, R., Wilde, B., Keiter, P., et al.: 2004, *Bull. Am. Phys. Soc.* **49** NO3 3, 255.

Coppins, M., Bond, D. J. and Haines, M. G.: 1984, *Phys. Fluids* **27**, 2886-2889.

DeSouza-Machado, S., Hassam, A. B. and Sina, R.: 2000, *Phys. Plasmas* **7**, 4632.

Fendt, C. and Čemeljić, M.: 2002, *A&A*, **395**, 1045.

Freidberg, J. P.: 1987, *Ideal Magnetohydrodynamics*, (Plenum, New York).

Golingo, R. P., Shumlak, U., and Nelson, B. A.: 2005, *Phys. Plasmas* **12**, 062505.

Keppens, R., Baty, H. and Casse, F.: 2005, *Space Sci. Rev.* **121**, 65.

Königl, A., and Pudritz, R. E.: 2000, *Protostars and Planets IV*, Edited by V. Mannings, A. P. Boss, and S. S. Russell, (The University of Arizona Press, Tucson).

Kruskal, M. D., and Schwarzschild, M.: 1956, *Proc. R. Soc. London, Ser. A* **223**, 348.

Lery, T. and Frank, A.: 2000, *ApJ*, **533** 897.

Lesch, H., Appl, S. and Camenzind, M.: 1989, *A&A* **225**, 341.

Mundt, R., Ray, T. P., Buhrke, T., Raga, Z. C., and Solf, J.: 1990, *A&A* **232**, 37.

Nakamura, M. and Meier, D., L.: 2004, *AIP Conference Proceedings*, **703**, 308.

Ray, T. Muxlow, T. W. B., Axon, D. J., Brown, A., Corcoran, D., Dyson, J. and Mundt, R.: 1997, *Nature* **385**, 415.

Ray, T.: 2004, *Bull. Amer. Phys. Soc.* **49**, 70.

Reipurth, B., and Heathcote, S.: 1997, *Herbig-Haro Flows and the Birth of Low Mass Stars*, Edited by B. Reipurth and C. Bertout, (IAU Symposium 182), p. 3.

Ruden, E.: 2002, *IEEE Trans. Plasma Sci.* **30**, 611.

Shafranov, V. D.: 1954, *At. Energ.* **5**, 38.

Shu, F. H., Najita, J. R., Shang, H., and Li, Z.: 2000, *Protostars and Planets IV*, Edited by V.

Shumlak, U., and Hartman, C. W.: 1995, *Phys. Rev. Lett.* **75**, 3285.

Sotnikov, V. I., Paraschiv, I., Makhin, V., Bauer, B. S., Leboeuf, J. N. and Dawson, J. M.: 2002, *Phys. Plasmas* **9**, 913.

Sotnikov, V. I., Bauer, B. S., Leboeuf, J. N., Helinger, P. Trávniček, P., and Fiala, V.: 2004, *Phys. Plasmas* **11**, 1897.

Spruit, H. C., Foglizzo, T., and Stehle, R.: 1997, *Mon. Not. R. Astron. Soc.* **288**, 333.

Vlemmings, W. H. T., Diamond, P. J. and Imai, J: 2006, *Nature* **440**, 58.

Völker, R., Smith, M. D., Suttner, G. and Yorke, H. W.: 1999, A&A **343**, 953.

Wanex, L. F., Sotnikov, V. I., Bauer, B. S., and Leboeuf, J. N.: 2004, *Phys. Plasmas.* **11**, 1372.

Wanex, L., F.: 2005a, *Astrophys. Space Sci.* **298**, 337

Wanex, L. F., Sotnikov, V. I., and Leboeuf, J. N.: 2005b, *Phys. Plasmas.* **12**, 042101.

Winterberg, F.: 1985, *Beitr. Plasmaphys.* **25**, 117.

Winterberg, F.: 1999, *Z. Naturforsch.* **54a**, 459.

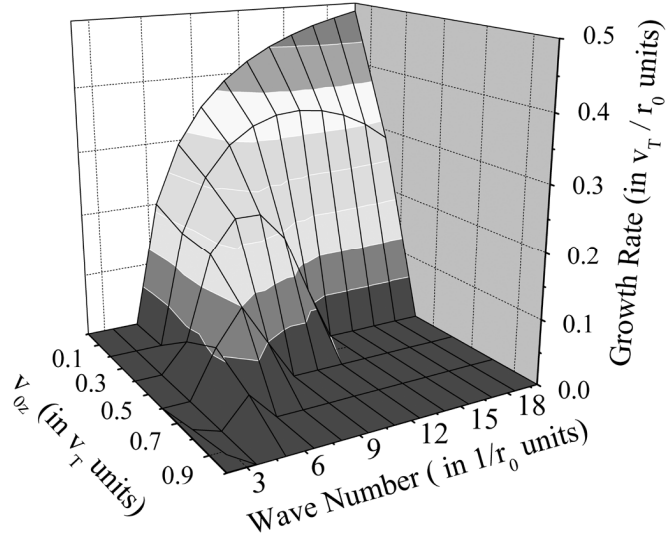

**Figure 1**. This is a 3D plot of the kink instability growth rates for the constant current density equilibrium profile with $v_{0\phi} = 0.1$ and $a = 0.1$. The growth rate is shown on the z-axis (in units of $v_T/r_0$), the wave number is shown of the y-axis (in units of $1/r_0$) and the value of the coefficient $v_{0z}$ on the x-axis (in units of $v_T$). As an example of how to interpret the plot consider the kink instability growth rate for $v_{0z} = 0.3$, the growth rate for axial wave numbers 2 and 3 are zero, the growth rate then increases with increasing wave number to a maximum of ~ 0.25 for wave numbers 8 and 9 and then decreases to zero for wave numbers above ~ 14. Instability growth rates are zero for wave numbers 2-20 for $v_{0z} > 1$.

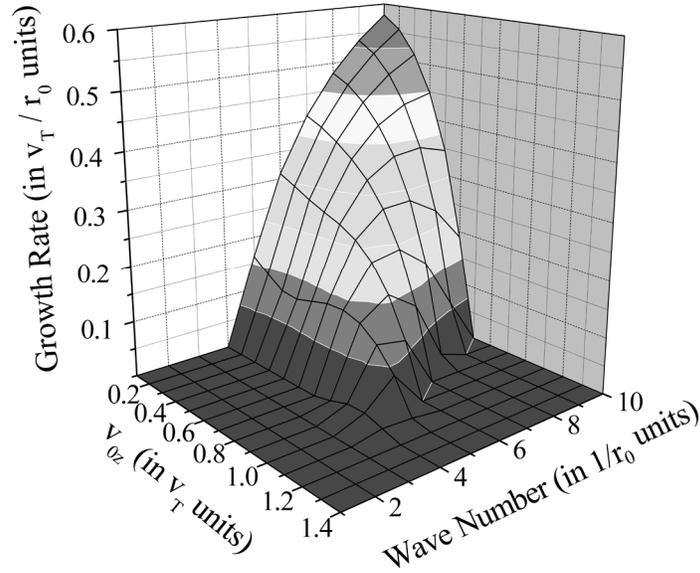

**Figure 2**. This is a 3D plot of the kink instability growth rates for the constant electron velocity equilibrium profile with $v_{0\phi} = 0.1$ and $a = 0.1$. The growth rate is shown on the z-axis (in units of $v_T/r_0$), the wave number is shown on the y-axis (in units of $1/r_0$) and the value of the coefficient $v_{0z}$ on the x-axis (in units of $v_T$). As an example of how to interpret the plot consider the kink instability growth rate for $v_{0z} = 0.9$, the growth rate for axial wave numbers 1 to 3 are zero, the growth rate then increases with increasing wave number to a maximum of ~ 0.1 for wave number 5 and then decreases to zero for wave numbers above ~ 7. Instability growth rates are zero for wave numbers 1-10 for $v_{0z} > 1.2$.